\documentclass[fleqn, usenatbib, useAMS]{mnras}
\usepackage[fleqn]{amsmath}
\usepackage[none]{hyphenat}
\usepackage{times}
\usepackage{graphicx}
\usepackage{color}
\usepackage{blindtext}
\title[Pre-impact orbits of giant impactors]{Constraints on the pre-impact orbits of Solar System giant impactors}
\author[A.P. Jackson et al.]{Alan P. Jackson$^{1,2}$\thanks{E-mail: ajackson@cita.utoronto.ca}, Travis S.J. Gabriel$^{2}$, Erik I. Asphaug$^{2,3}$\\%
$^1$Centre for Planetary Sciences, University of Toronto, 1265 Military Trail, Toronto, Ontario, M1C 1A4, Canada\\%
$^2$School of Earth and Space Exploration, Arizona State University, 781 E. Terrace Mall, Tempe, Arizona 85287, USA\\%
$^3$Lunar and Planetary Laboratory, University of Arizona, 1269 E. University Blvd, Tucson, Arizona 85721, USA}
\date{Submitted 2017}
\pagerange{\pageref{firstpage}--\pageref{lastpage}}
\pubyear{2017}
\begin{document}
 \label{firstpage}
 \maketitle
 
 \begin{abstract}
  We provide a fast method for computing constraints on impactor pre-impact orbits, applying this to the late giant impacts in the Solar System.  These constraints can be used to make quick, broad comparisons of different collision scenarios, identifying some immediately as low-probability events, and narrowing the parameter space in which to target follow-up studies with expensive $N$-body simulations.  We benchmark our parameter space predictions, finding good agreement with existing $N$-body studies for the Moon.  We suggest that high-velocity impact scenarios in the inner Solar System, including all currently proposed single impact scenarios for the formation of Mercury, should be disfavoured.  This leaves a multiple hit-and-run scenario as the most probable currently proposed for the formation of Mercury.
 \end{abstract}

 \begin{keywords}
  planets and satellites: formation, planets and satellites: terrestrial planets, methods: numerical, celestial mechanics, Moon
 \end{keywords}

 \section{Introduction}
 \label{sec:intro}
 
 Giant, planetary scale impacts are a key component of the final phase of terrestrial planet formation \citep[e.g.][]{kenyon2006, raymond2009, kokubo2010}.  Within our own Solar System evidence for giant impacts is ubiquitous.  A giant impact event is the leading theory for the formation of Earth's Moon \citep[e.g.][]{cameron1976, canup2004b}, for which the impactor has been named `Theia' by \citet{halliday2000}.  On Mars a massive impact appears to be the best explanation for the hemispheric dichotomy \citep[e.g.][]{wilhelms1984, marinova2008, nimmo2008, marinova2011}, such that the northern lowlands are then rather the hemisphere-scale Borealis impact basin.  The large core fraction of Mercury has also been suggested to have resulted from formation in a giant impact event \citep{anic2006, benz2007, asphaug2014b}.  Indeed, the only terrestrial planet for which there is no clear evidence of a giant impact event is Venus, and it is likely no coincidence that this is also the terrestrial planet whose surface and geological history are least well understood.  Giant impacts are also not limited to the inner Solar System; giant impact scenarios are proposed, for example, for the Pluto-Charon system \citep[e.g.][]{stern2006, canup2011} and Haumea \citep{leinhardt2010}.
 
 Each of the giant impact events identified above is postulated as being the final one experienced by the body in question - there may have been earlier events, and indeed for Earth it is likely there were, but direct evidence for any earlier giant impact events was erased by the last.  As these events played such an important role in determining the final basic characteristics (e.g. composition, angular momentum, state of differentiation) of the target planet, and the formation of satellites and dynamical relationships with remnants, if we want to understand the history of the Solar System it is important to examine where the impactors might have originated.  
 
 For the Moon in particular there are concerns about matching the isotopic similarity with Earth given the dynamical constraints and the physics of collisions \citep[e.g.][]{asphaug2014}.  Meteorites display a diversity of isotopic signatures widely believed to be the result of chemical gradients and inhomogeneneities in the Solar nebula \citep[e.g.][]{clayton2003a}.  While there is evidence of a decrease in isotopic diversity for larger bodies \citep[e.g.][]{ozima2007}, presumably due to accretional mixing across reservoirs, differences are quite substantial between Mars and other achondrite parent bodies. It is therefore expected that giant impactors and their targets will have readily discernible isotopic signatures \citep{kaib2015} that will be evident in the compositions of the final planets and their satellites.  Outlining the origin of giant impactors can thus provide testable geochemical constraints and help validate models for interpreting isotope geochemistry \citep[e.g.][]{young2016}.
 
 Because giant impacts mix and disrupt the compositions of their contributing materials, a consistent and correct geochemical interpretation requires understanding the provenance of the projectile. In some models for the impact origin of the Moon, for example, Theia is a fast interloping protoplanet from the outer solar system, that collided at several times the escape velocity $v_{\rm esc}$, while in other models it is a Mars-like planet that accreted nearby and collided just barely over $v_{\rm esc}$. Dynamical analysis is therefore a required compliment to geochemical analysis, and the parameters of a giant impact event can provide unique insight into the orbit of the impactor immediately prior to the collision, allowing one to assess the likelihood of the projectile deriving from a given source region in the solar system. 
 
 \citet{rivera2002} and \citet{quarles2015} explored the possible origins of Theia using suites of N-body integrations, to find locations in which a potential Theia can remain quasi-stable for several tens of Myr before colliding with Earth to produce a system reminiscent of the present terrestrial planets.  While N-body simulations afford perhaps the most detailed method of examining the origins and timing of an impact event, they are computationally expensive and thus one is inevitably forced to limit the parameter space under investigation. The motivation for this work is to obtain a more general analytical basis for examining giant impact projectile origins, that can be applied to the several competing hypotheses for Moon formation, as well as to giant impact scenarios involving Mercury, Mars, Pluto and other planets.
 
 The impact velocity $v_{\rm imp}$ is one of the most important parameters in determining the outcome of a planet-forming collision \citep[e.g.][]{asphaug2010, leinhardt2012}. In terms of the pre-impact orbits, $v_{\rm imp}$ is equal to the relative velocity between the orbits of the target and impactor at the point of impact.  If we can constrain the orbit of the target, then we can then use this information to place constraints on the pre-impact orbit of the impactor, and hence, its provenance and possible composition. In most giant impacts the projectile is much less massive than the target, in which case we expect the orbit of the target to be close to the orbit of the final planet as we see it today. In making this assumption, the allowable projectile orbits can be computed analytically much faster than using N-body simulations, enabling the exploration of much larger parameter spaces and a wider range of giant impact scenarios.
 
 In this work we analyse a variety of scenarios for the last several giant impacts believed to have occurred in the Solar System. We use the velocities of those collisions as interpreted by various published models, to constrain the pre-impact orbits of the impactor bodies, e.g. Theia.  First, in Section~\ref{sec:method}, we outline our methods for relating the pre-impact orbits to the orbit of the target body.  In Section~\ref{sec:moon} we then apply this to the Moon-forming giant impact for several different impact scenarios; the `Canonical' model (\citealt{hartmann1975, cameron1976}, as reviewed by e.g. \citealt{canup2004b}), the hit-and-run model of \citet{reufer2012}, the equal-mass impactors model of \citet{canup2012}, and the small, fast impactor model of \citet{cuk2012}, before comparing to the results of \citet{quarles2015} as a benchmark.  We then apply our pre-impact orbit determination schema to the formation of Mercury \citep{benz2007, asphaug2014b} in Section~\ref{sec:mercury}.  In Section~\ref{sec:borealis} we briefly discuss the Borealis basin impact on Mars \citep{marinova2008, marinova2011} as an example of a scenario where the target has an eccentric orbit, and other Solar System giant impacts including Pluto-Charon in Section~\ref{sec:otherimpacts}.  Finally, we bring together our conclusions in Section~\ref{sec:conclusions}. 
 
 \section{Method}
 \label{sec:method}
 
 \citet{jackson2014} (hereafter \defcitealias{jackson2014}{J14}\citetalias{jackson2014}) presented equations relating the orbit of a particle before and after it receives a `kick' to its velocity.  They used these in the context of relating the orbits of debris produced in an impact event to the orbit of the progenitor body.  We can however reverse the causal relationship and use the same equations to relate the orbit of an impactor to that of the target body and the relative velocity between the two at the time of impact.
 
We direct the reader to Section~2 of \citetalias{jackson2014} for the full derivation of these equations, and a detailed discussion of their behaviour, however for reference we reproduce the equations we shall use here, for circular target orbits:
\begin{equation}
\label{aaprim}
\frac{a_{\rm T}}{a_{\rm I}}=1-2\left(\frac{v_{\rm rel}}{v_{\rm k}}\right) S_{\theta}S_{\phi}-\left(\frac{v_{\rm rel}}{v_{\rm k}} \right)^2,
\end{equation}
\begin{equation}
\label{eprim2}
e_{\rm I}^2=1-\left(\frac{h_{\rm I}^2}{h_{\rm T}^2}\frac{a_{\rm T}}{a_{\rm I}}\right),
\end{equation}
\begin{equation}
\label{Iprim}
\cos I_{\rm I}=\left[1+\left(\frac{v_{\rm rel}}{v_{\rm k}}\right)S_{\theta}S_{\phi}\right]\left(\frac{h_{\rm I}^2}{h_{\rm T}^2}\right)^{-\frac{1}{2}},
\end{equation}
\begin{equation}
\label{hhprim}
\frac{h_{\rm I}^2}{h_{\rm T}^2}=1+2\left(\frac{v_{\rm rel}}{v_{\rm k}}\right)S_{\theta}S_{\phi}+\left(\frac{v_{\rm rel}}{v_{\rm k}}\right)^2(C_{\theta}^2+S_{\theta}^2S_{\phi}^2),
\end{equation}
and for eccentric target orbits:
\begin{equation}
\label{aaprimecc}
\frac{a_{\rm T}}{a_{\rm I}}=1 - \left(\frac{v_{\rm rel}}{v_{\rm k}}\right)^2 - \frac{2}{\sqrt{1-e_{\rm T}^2}}\left(\frac{v_{\rm rel}}{v_{\rm k}}\right) S_{\theta}(S_{(\phi-f_{\rm T})} + e_{\rm T} S_{\phi}),
\end{equation}
\begin{equation}
\label{ecceprim}
e_{\rm I}^2=1-(1-e_{\rm T}^2)\left(\frac{h_{\rm I}^2}{h_{\rm T}^2}\frac{a_{\rm T}}{a_{\rm I}}\right),
\end{equation}
\begin{equation}
\label{eccIprim}
\cos I_{\rm I}=\left[1+\frac{(1-e_{\rm T}^2)^{\frac{1}{2}}}{1+e_{\rm T} C_{f_{\rm T}}} \left(\frac{v_{\rm rel}}{v_{\rm k}}\right)S_{\theta}S_{(\phi-f_{\rm T})}\right]
              \left(\frac{h_{\rm I}^2}{h_{\rm T}^2}\right)^{-\frac{1}{2}},
\end{equation}
\begin{equation}
\label{ecchprimh}
\begin{split}
\frac{h_{\rm I}^2}{h_{\rm T}^2} =1 &+ 2 \frac{(1-e_{\rm T}^2)^\frac{1}{2}}{1+e_{\rm T} C_{f_{\rm T}}} \left(\frac{v_{\rm rel}}{v_{\rm k}}\right) S_{\theta} S_{(\phi-f_{\rm T})} \\
& +\frac{1-e_{\rm T}^2}{(1+e_{\rm T} C_{f_{\rm T}})^2} \left(\frac{v_{\rm rel}}{v_{\rm k}}\right)^2 \left(C_{\theta}^2 + S_{\theta}^2 S_{(\phi-f_{\rm T})}^2\right).
\end{split}
\end{equation}
We make some minor alterations to the notation of \citetalias{jackson2014} to emphasise the purpose for which we are using the equations here.  The subscripts $\rm T$ and $\rm I$ refer to orbital elements and quantities of the target and of the impactor respectively.  The velocity kick between the orbits is denoted as $v_{\rm rel}$, since it is the relative velocity at the point of impact in this case.  Otherwise we follow the same notation as \citetalias{jackson2014}:
\begin{tabular}{r p{0.7\columnwidth}}
 $a$ & semi-major axis, \\
 $e$ & eccentricity, \\
 $I$ & inclination, \\
 $f$ & true anomaly, \\ 
 $\Omega$ & longitude of ascending node, \\
 $\omega$ & argument of pericentre, \\
 $v_{\rm k}$ & circular speed at orbital distance $a$, \\
 $S_x$ & $\sin(x)$, \\
 $C_x$ & $\cos(x)$, \\
 $\theta$ and $\phi$ & spherical polar coordinate angles defining the orientation of the relative velocity. \\
\end{tabular}
Since orbits precess over relatively short timescales the orientation of the orbits, $\Omega$ and $\omega$ are not of interest here and the equations for $\Omega_I$ and $\omega_I$ are omitted.  Similarly the true anomalies at the time of impact are not of interest other than in the true anomaly of the target defining the range of possible impactor orbits.  Note also that since $v_{\rm rel}$ is defined at the instant of impact the impulsive assumption of \citetalias{jackson2014} is always satisfied.

For a given target orbit and relative velocity there is a unique mapping between $\theta$ and $\phi$ and the impactor orbit, however $\theta$ and $\phi$ are not known so we assume them to be isotropically distributed.  As such each target orbit and relative velocity maps to a range of possible impactor orbits.

\section{The Moon}
\label{sec:moon}
 
The formation of the Moon is the most well-studied giant impact event in the Solar system.  The modern theory was first developed in the 1970s, following the Apollo missions, which revealed a completely igneous planet with very little metallic iron \citep[e.g.][]{hartmann1975, cameron1976}, and over decades of refinement produced what has come to be known as the `Canonical' model for Moon-formation \citep[e.g.][]{canup2004b}.  In this model a roughly Mars-sized Theia strikes the nearly fully-formed proto-Earth at around $45^{\circ}$ (the most probable angle) at just above the mutual escape velocity, in what can be described as a `graze-and-merge' collision, with most of Theia ultimately accreted by Earth.

The Canonical model can well re-produce many aspects of the Earth-Moon system, including mass, angular momentum and the depletion of iron in the Moon relative to Earth.  However, a large fraction of the material that forms the Moon in the Canonical model is derived from Theia, whereas the growing body of isotopic evidence (especially oxygen) suggests that the Moon is nearly identical to Earth, leading to a conflict with the Canonical model \citep[e.g.][]{asphaug2014}.  This conflict has initiated many new studies to investigate how it might be resolved, including a renewed interest in alternative impact scenarios.  As such we will analyse a total of 4 models for the Moon-forming giant impact: the Canonical model, and those of \citet{reufer2012}, \citet{canup2012}, and \citet{cuk2012}.  Throughout we assume that the target body occupies a circular orbit at 1~AU and inclinations are measured relative to the plane of this orbit.
 
\subsection{Canonical and \citet{canup2012} models}
\label{sec:moon:canonical}
 
The Canonical model involves an impact between a proto-Earth of around 0.9~$M_{\oplus}$ and a Mars-sized Theia of around 0.1~$M_{\oplus}$ colliding at an oblique angle of around 45$^{\circ}$ at a low speed of around 1.05~$v_{\rm esc}$, where $v_{\rm esc}$ is the mutual escape velocity \citep[e.g.][]{canup2001, canup2004a, canup2004b}. The model is Canonical in that it takes the angular momentum of the final Earth-Moon system as a conserved quantity, that has to be close to the value observed today. According to \cite{cuk2012} the original angular momentum could have been more than twice as great, leading to diverse new models. \citet{canup2012} suggest an impact between two nearly equal mass semi-Earths at a lower angle, but with a very similar velocity.  Both the Canonical model and the semi-Earths model use very similar relative velocities, and will therefore produce similar results for the potential pre-impact orbits of Theia, albeit that the assumption that the orbit of the target is that of the present day Earth is probably less accurate for the model of \citet{canup2012}.
 
\begin{figure}
 \includegraphics[width=\columnwidth]{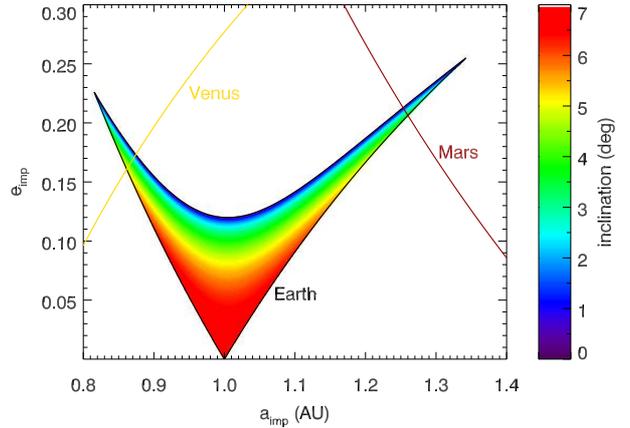}
 \caption{The allowed semi-major axis -- eccentricity -- inclination space for Theia in the Canonical and \citet{canup2012} models for an impact velocity of 1.05~$v_{\rm esc}$.  Orbits above the yellow line are Venus-crossing, while those above the dark red line are Mars-crossing.}
 \label{fig:EMcanonicalaei}
\end{figure}

In Fig.~\ref{fig:EMcanonicalaei} we show the range of pre-impact orbital parameters allowed for Theia under the Canonical and \citet{canup2012} models.  In the 3 dimensions of semi-major axis, eccentricity and inclination this is a thin surface, which we represent here in 2D by applying a colour map to represent the inclination.  The impact velocity of 1.05~$v_{\rm esc}$ corresponds to $v_{\rm rel}/v_{\rm k}$=0.12, or $v_{\rm rel}$=3.5~km~s$^{-1}$.  This is a very low impact velocity, only just above the 1~$v_{\rm esc}$ of a pure free-fall impact and hence the range of pre-impact orbits is also quite small.

The 3D surface of allowed pre-impact orbits of Theia occupies a distinctive `V' shape when projected into semi-major axis and eccentricity which is characteristic of this type of problem.  The inner (low semi-major axis) boundary corresponds to potential Theian orbits that impact Earth when Theia is at apocentre, while the outer (high semi-major axis) boundary corresponds to potential Theian orbits that impact Earth when Theia is at pericentre.  The upper boundary is the line along which $I_{\rm imp}$=0$^{\circ}$. In yellow and dark red we show the lines above which Theia will cross the orbits of Venus and Mars respectively.  As seen, in these very low relative velocity scenarios, the ranges of allowed orbits in the Venus and Mars crossing regions are small and most of the allowed pre-impact orbits for Theia only cross the orbit of Earth.
 
 While we only directly constrain the orbit of Theia immediately before the impact, the allowed orbital parameters suggest that in these two scenarios Theia likely originated from near the proto-Earth.  It is relatively easy to excite the eccentricity of an early solar system body over millions of years, and so we would expect that Theia could start off somewhat below the coloured region in Fig.~\ref{fig:EMcanonicalaei} before being excited up into the allowed pre-impact region later.  Indeed this would fit well with the Moon-forming impact occurring somewhat late in the process of Solar System formation.  
 
 Dynamical pathways that involve bringing Theia from a more distant birthplace (in semi-major axis) onto the pre-impact orbit are more difficult to construct.  Such pathways necessarily involve scattering by other Solar System planets, and as we can see in Fig.~\ref{fig:EMcanonicalaei}, the regions of overlap with Venus or Mars crossing orbits are small, so that any such multiple scattering pathway would have to be rather fine-tuned to produce the correct relative velocity at impact.  This expectation that Theia originates from within roughly the semi-major axis range allowed for the pre-impact orbit is supported by published detailed dynamical models, as we discuss in Section~\ref{sec:moon:quarles}.

If, by the time of final accretion, there was a constant gradient in isotopic composition with distance from the Sun, this would imply that the difference in compositions between proto-Earth and a nearby Theia were less than that between Earth and Mars or Earth and Venus. However the isotopic dissimilarity between the Earth and Mars is pronounced, and the Moon predicted by the Canonical model, being made mostly out of Theia material, would still be impossible to explain without Theia forming out of almost identical materials as the Earth, or else a vigorous, whole-mantle mixing mechanism to fully or partially homogenise isotopic abundances between the post-impact Earth and proto-lunar disk \citep[e.g.][]{pahlevan2007}, or some form of post-formation processing \citep{salmon2012}. The near-identity of Earth-Moon isotopic abundances \citep[e.g.][]{young2016} is a critical point to bear in mind when comparing the different scenarios.
 
\subsection{\citet{reufer2012} model}
\label{sec:moon:reufer}
 
\citet{reufer2012} proposed an alternative giant impact scenario with a `hit-and-run' impact in which a substantial silicate-dominated fraction of Theia continues downrange after the impact and is not incorporated into the Earth-Moon system.  Hit-and-run requires a slightly higher impact velocity than the Canonical scenario, of 1.2~$v_{\rm esc}$, which corresponds to $v_{\rm rel}/v_{\rm k}$=0.25, or $v_{\rm rel}$=7.4~km~s$^{-1}$, about twice the relative velocity of the Canonical scenario.  Thus it brings more impact energy, but leaves behind less Theia-derived material, than the Canonical scenario. Because only a fraction of the proto-lunar disk derives from Theia in the hit-and-run scenario, it remains isotopically more similar to Earth. Forming from a hotter disk, the protolunar materials then have potential for further isotopic equilibration. 
 
\begin{figure}
 \includegraphics[width=\columnwidth]{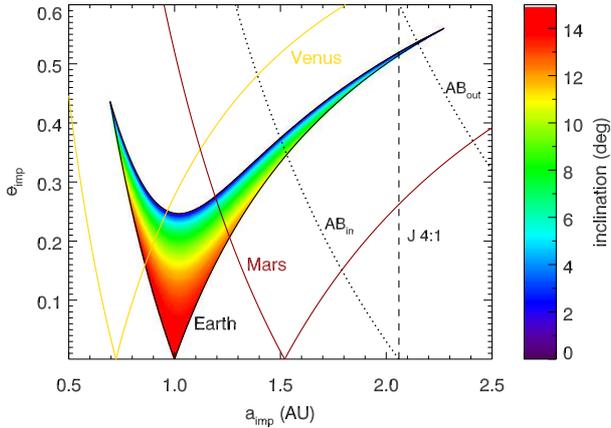}
 \caption{The allowed semi-major axis -- eccentricity -- inclination space for Theia in the \citet{reufer2012} models for an impact velocity of 1.2~$v_{\rm esc}$.  Orbits above the yellow line are Venus-crossing, while those above the dark red line are Mars-crossing.  The dashed line indicates the 4:1 mean-motion resonance with Jupiter at 2.06~AU, while the dotted lines indicate orbits that have apocentres at 2.06 and 3.3~AU, the inner and outer edges of the main asteroid belt.}
 \label{fig:EMreuferaei}
\end{figure}
 
As we show in Fig.~\ref{fig:EMreuferaei}, owing to the increased velocity, the allowed range of orbital parameters for the pre-impact orbit of Theia is significantly expanded in the hit-and-run scenario as compared with the Canonical or \citet{canup2012} scenarios.  As a result a significant portion of the parameter space is now in regions that will cross the orbits of Venus or Mars, in addition to the proto-Earth.  In particular there is a tail that extends somewhat past the 4:1 mean-motion resonance with Jupiter at 2.06~AU that marks the approximate inner edge of the asteroid belt.  For considering potential origins of Theia in the region of the present asteroid belt, a more relevant comparison is the apocentre of the pre-impact orbit, since Theia will have been perturbed from its initially quasi-stable orbit before impact.  The region between the two dotted lines thus indicates those orbits which have apocentres between 2.06 and 3.3~AU, which encompasses the Main Belt.

The possibility that Theia originated in the Main Belt, indeed around the orbit of Ceres, is particularly relevant to composition, because the region $\sim 2-3$ AU and beyond was considerably more volatile-rich than around 1 AU during planet formation. One of the collisional scenarios that \citet{reufer2012} identify as a good fit to producing the Earth-Moon system, is one in which the impactor is water-rich, as might be expected if Theia originated from the outer Main Belt. Ceres, the asteroid at $a=2.8$ AU that contains almost half the mass of the present Main Belt, has a substantial mass fraction of water \citep{thomas2005,russell2016dawn}, and more massive planetesimals, since lost from the region between Mars and Jupiter, might have had even greater water fractions \citep[e.g.][]{asphaug2017planetesimals}. At 0.15-0.2~$M_{\oplus}$ Theia might have represented an anomalously large Main Belt planetesimal by the time of the presumed Moon-forming giant impact $\sim$50-150~Myr after the first solids. Still, simulations have shown that individual large bodies can survive in the vicinity of the asteroid belt for up to $\sim$1~Gyr before colliding with a terrestrial planet \citep[e.g.][]{chambers2007}.

\subsection{\citet{cuk2012} model}
\label{sec:moon:cuk}

Whereas the scenarios of \citet{reufer2012} represent only a relatively minor departure from the Canonical model, at least in terms of the relative velocity required, the scenarios proposed by \citet{cuk2012} are a much more dramatic deviation. In their scenario the proto-Earth is already a rapidly rotating object, with spin period $<3$ hrs, and Theia is a smaller body that strikes at much higher velocity, retrograde to the Earth's spin. Arguing that final Earth-Moon angular momentum could be twice that of the present day, \citet{cuk2012} introduce models that successfully reproduce other aspects of the Earth-Moon system. Required impact velocities are 2-2.5~$v_{\rm esc}$, corresponding to $v_{\rm rel}/v_{\rm k}$=0.65-0.86, or $v_{\rm rel}$=19.3-25.6~km~s$^{-1}$, almost a six-fold increase from the relative velocity of the Canonical scenario.

\begin{figure}
 \includegraphics[width=\columnwidth]{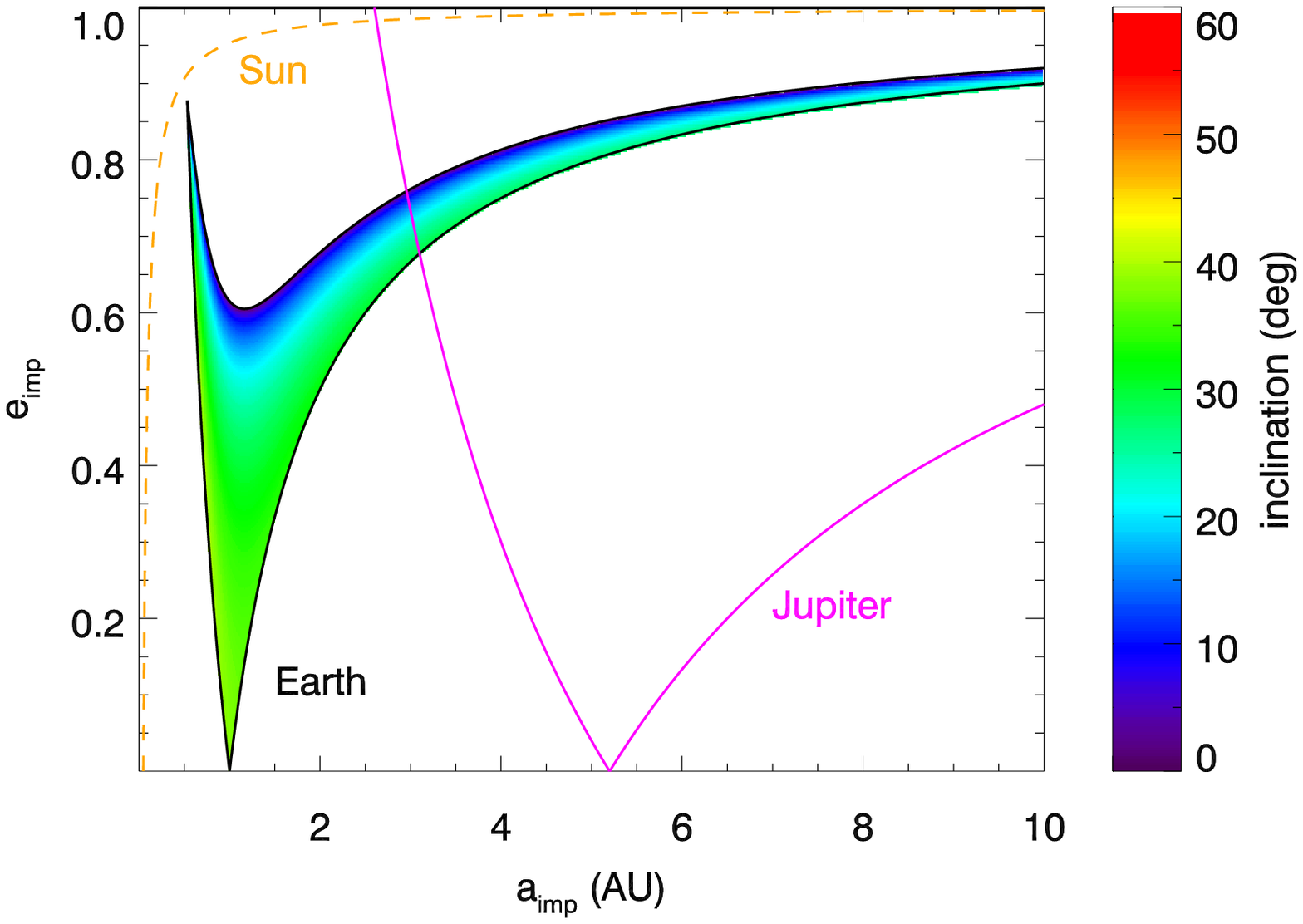}
 \includegraphics[width=\columnwidth]{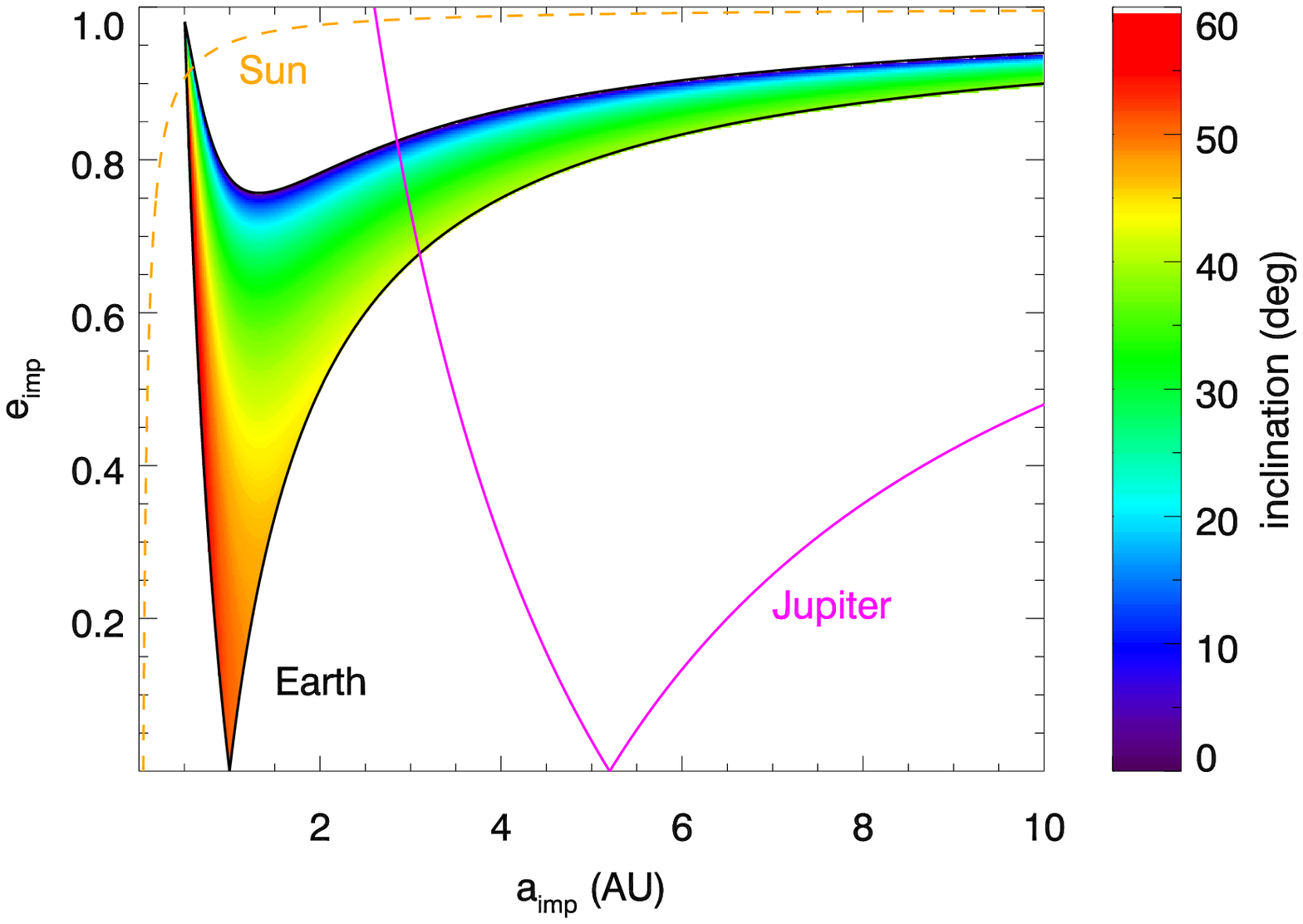}
 \caption{Semi-major axis -- eccentricity -- inclination space for the pre-impact orbit of Theia in the \citet{cuk2012} models for an impact velocity of 2.0$v_{\rm esc}$ (top) and 2.5$v_{\rm esc}$ (bottom).  Orbits above the magenta line are Jupiter crossing, while those above the dashed orange line are Sun grazing (having pericentres within 10~$R_{\odot}$ of the Solar surface).}
 \label{fig:EMCSaei}
\end{figure}

As would be expected given the much larger relative velocities required by this scenario, the parameter space of pre-impact orbits shown in Fig.~\ref{fig:EMCSaei} is much larger than that in Figs.~\ref{fig:EMcanonicalaei} and \ref{fig:EMreuferaei}.  These very high relative velocities actually allow for hyperbolic orbits, such that it is not possible to show the entire parameter space in Fig.~\ref{fig:EMCSaei}.  While the allowed parameter space based solely on the relative velocity is very large, we can use additional constraints to reduce the plausible range.  Bodies that pass extremely close to the Sun will experience extreme heating and strong tides that can alter the structure and orbit of the body.  As such impactors on Sun-grazing orbits are implausible.  We also note that there is today a complete lack of material interior to Mercury and that terrestrial planet formation models also do not generally include material so close to the Sun \citep[e.g.][]{raymond2009,chambers2013}.  The Sun-grazing region does not have a sharp edge, but as a representative boundary in Fig.~\ref{fig:EMCSaei} we show in orange the line above which Theia comes within 10~$R_{\odot}$ at pericentre.

We also acknowledge the model of \citet{rufu2017}, where the Moon is generated through the accretion of several smaller moonlets, each moonlet generated in a series of lesser (and individually more probable) giant impacts into a rotating proto-Earth. Each subsequent impact would either diminish or enhance the overall angular momentum of the system. Across their simulations of giant impacts with the target rotating at $\omega=0.1, 0.25, 0.5 \omega_{\rm{max}}$, where $\omega_{\rm{max}}$ is the spin-limit of the target, the scenarios that generate debris discs composed predominantly of Earth material require impact velocities of $v_{\rm{imp}} \geq 2.0 v_{\rm{esc}}$. So in terms of required velocity, we view the model of \citet{rufu2017} in the same light as that of \citet{cuk2012}, but instead of requiring a single, high-velocity impact into a fast-rotating Earth, \citet{rufu2017} requires a series of impactors with high relative velocity.

\defcitealias{quarles2015}{QL15}
\begin{figure*}
 \includegraphics[width=\textwidth]{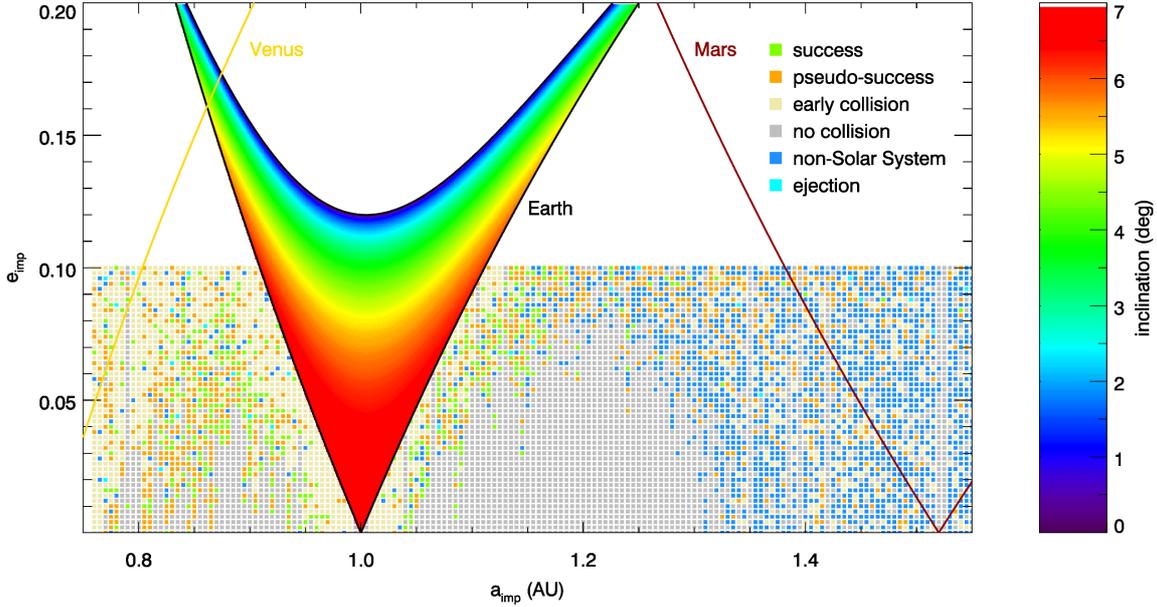}
 \caption{Our semi-major axis -- eccentricity -- inclination distribution for the Canonical (and \citealt{canup2012}) scenario overlaid on the results of \citetalias{quarles2015} as shown in their Fig. 2d for the Nice4 dataset.  The key at top right indicates the different collision outcomes in the background plot, see \citetalias{quarles2015} for the full definitions.  Data from \citetalias{quarles2015} provided by B. Quarles.}
 \label{fig:EMcanon-QL}
\end{figure*}

\subsection{Comparison to \citet{quarles2015}}
\label{sec:moon:quarles}

The analytical approach developed here can be compared directly with numerical simulations, in the case of Earth-Moon system formation. \citet{quarles2015} (hereafter \citetalias{quarles2015}) conducted a detailed set of $N$-body integrations of a model early solar system with 5 terrestrial planets: Mercury, Venus, Mars, the proto-Earth and Theia, looking for successful cases in which Theia collides with the proto-Earth between 8 and 200~Myr after the start of the simulation, the widest possible range of ages for Moon formation. The majority of their simulations (their Table 3) begin with the semi-major axis of Theia lying between 0.76 and 1.55~AU, covering the range allowed for both the Canonical and semi-Earths \citep{canup2012} scenarios.  They conduct a smaller set of simulations extending the range of initial semi-major axes for Theia to 0.44 and 2.18~AU, thereby also largely covering the range allowed for the hit-and-run \citep{reufer2012} scenario. 

In most successful cases of Moon formation, \citetalias{quarles2015} find that Theia originates from inside 1.3~AU, and that from 0.76-1.3~AU the distribution of initial semi-major axes for Theia is fairly flat (their Fig. 5).  Origins for Theia outside 1.3~AU are much more likely to result in an outcome that does not look like the Solar System. In Fig.~\ref{fig:EMcanon-QL} we overlay our allowed pre-impact semi-major axis -- eccentricity -- inclination distribution for the Canonical and \citet{canup2012} scenarios (as shown in Fig.~\ref{fig:EMcanonicalaei}) on the distribution of simulation outcomes for the Nice4 dataset of \citetalias{quarles2015} (after their Fig. 2d).  We can see that the initial conditions for which they identify success or pseudo-success cluster just outside our allowed pre-impact orbit distribution. This is expected as it allows Theia to have a substantial period of quasi-stability before moving onto a crossing orbit as defined by the bounded area in Fig.~\ref{fig:EMcanon-QL} and impacting the proto-Earth.

For their broader set of simulations that allows initial semi-major axes for Theia of 0.44-2.18~AU \citetalias{quarles2015} divide the results into 3 broad bins by initial semi-major axis, $0.44\leq a\leq 0.75$, $0.75< a\leq 1.55$, and $1.55< a\leq 2.18$.  Within the inner bin they find that 4.7\% of collisions are successful, in the middle bin 14\% are successful, and in the outer bin 6.3\% are successful.  A further 11.4\%, 22\% and 20.9\% respectively fall into the pseudo-success category in which between 8 and 200~Myr after the start of the simulation there is a collision between any small body (Mercury, Mars, Theia) and any large body (Earth, Venus) other than Theia-Earth.

The middle bin corresponds roughly to the allowed semi-major axis range for the Canonical and \citet{canup2012} models, while the inner bin is largely interior to the allowed ranges for all three low-velocity models (Canonical, \citet{canup2012} and \citet{reufer2012}) and the outer bin covers the outer part of the allowed range in the \citet{reufer2012} model.  The inner bin has the lowest rate of successful impacts and thus matches with our expectations. While it is more difficult to examine the distribution in any single bin due to the lower number of simulations, all of the successful cases of the inner bin are concentrated at the larger semi-major axes.  The results of \citetalias{quarles2015} are thus consistent with our suggestion that $N$-body simulations to conduct detailed analysis of proposed giant impact scenarios should be concentrated in the regions indicated by the distributions for the pre-impact orbits.

As we mentioned in Section~\ref{sec:moon:canonical}, we could envisage dynamical pathways involving multiple scattering events with the other terrestrial planets, and/or secular orbital evolution to substantially change the semi-major axis of Theia prior to it being put onto the final impact trajectory.  In the particular cases of the Canonical and \citet{canup2012} scenarios this seems rather unlikely due to the small size of the overlap between the allowed pre-impact orbits and the Mars and Venus crossing regions.  The other scenarios have larger overlaps, but nonetheless any pathway involving multiple scattering events or additional orbital evolution is inherently more complex than one involving only a single scattering event, and so we would expect it to be less likely, unless the single scattering event is extreme such that it has a much lower probability.  This is supported by the $N$-body studies of \citetalias{quarles2015}, since they do indeed find that in successful impacts Theia originates from close to the distribution of pre-impact orbit parameters.

\subsection{Additional Constraints on the pre-impact orbit}
\label{sec:moon:addconstraints}

\begin{figure*}
    \centering
    \includegraphics[width=\textwidth]{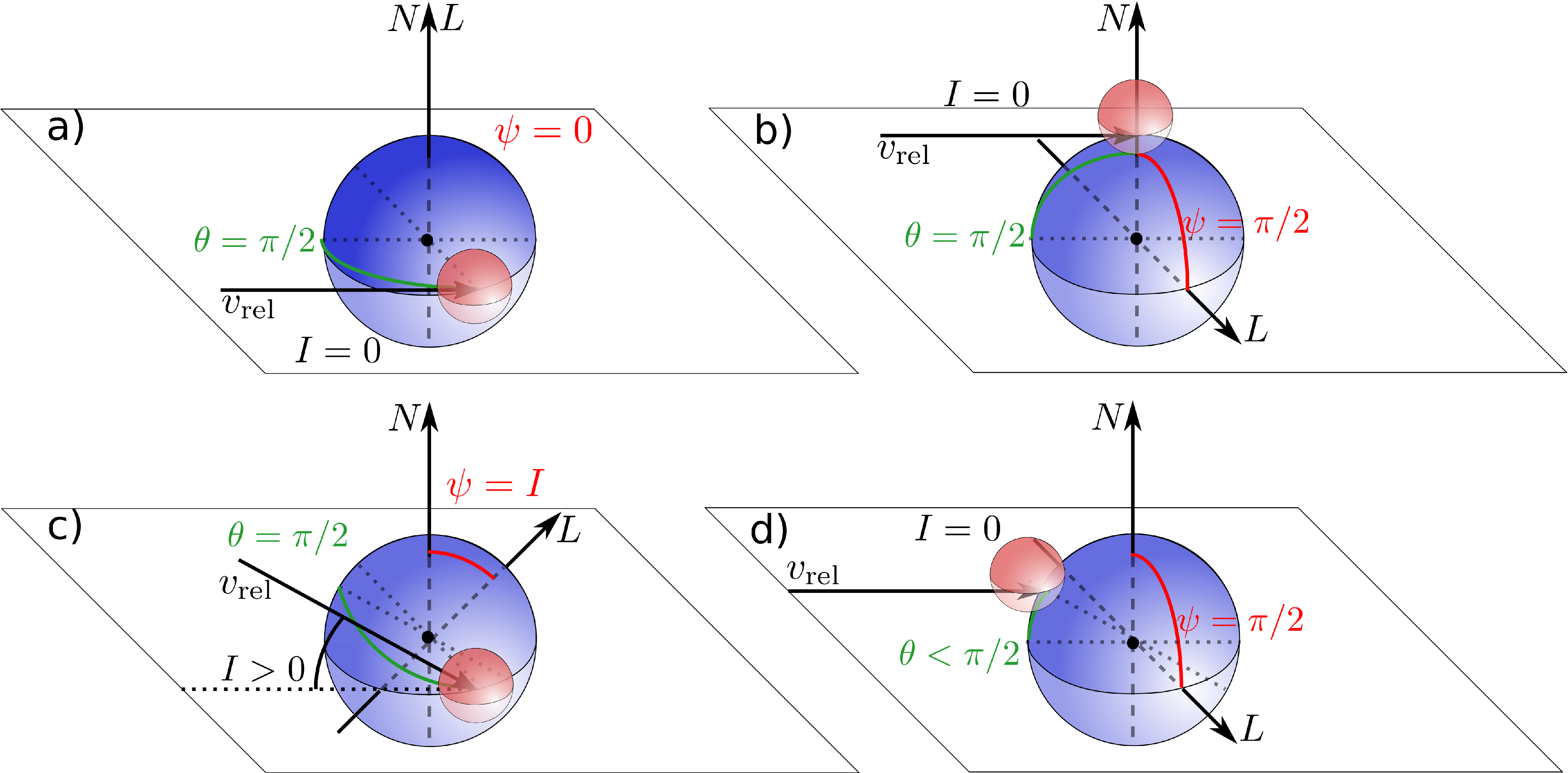}
    \caption{Diagram to illustrate the relationship between the orientation of the impact, the inclination and the obliquity.  The white plane represents the orbital plane of the target (the larger blue sphere), with the vector $N$ being the normal to the orbital plane.  The vector $L$ is the (post-impact) spin angular momentum of the target, with the pre-impact spin assumed to be zero.  The angle $\theta$ is the impact angle, measured between the centres of the bodies at the point of impact and a line through the centre of the target parallel to the impact velocity vector, $\theta=\pi/2$ for a grazing impact and $\theta=0$ for a head-on impact.  The angle $\psi$ is the obliquity, the angle between the spin axis and the orbit normal.  The angle $I$ is the inclination, the angle between the orbital planes of the target and the impactor.  In panels a) and b) we illustrate how the same impact angle and inclination can result in very different obliquities if the target is struck at the `side' as opposed to the `top' or the `bottom'.  In panel c) we then show how for non-zero inclinations the minimum obliquity is greater than zero even if the target is struck at the side.  Finally panel d) shows an impact with the same orientation as b) but with a smaller impact angle.  While the magnitude of the spin angular momentum imparted may be smaller in this case the obliquity is the same.}
    \label{fig:obliquity}
\end{figure*}

In addition to the relative velocity of the impact, it is valuable to consider other quantities to further constrain the parameter space of allowed pre-impact orbits.  One such additional constraint comes from the obliquity of Earth and the inclination of the lunar orbit. Together these define the orientation of the total angular momentum vector of the Earth-Moon system.  

In all of the scenarios we have discussed above, with the exception of \citet{cuk2012}, the majority of the angular momentum of the post-impact Earth-Moon system is imparted by the giant impact (see \citealt{canup2008} for an explicit examination of the effect of pre-impact rotation for the Canonical scenario).  While work by \citet{cuk2012, wisdom2015} has suggested that Solar tides may be able to change the magnitude of the total angular momentum of the Earth-Moon system more than previously thought, the orientation of the total angular momentum vector is nonetheless still an important constraint.  Even for the scenarios of \citet{cuk2012}, though much of the final angular momentum of the system is derived from the rapid spin of the pre-impact proto-Earth, the orientation of the impact relative to the spin axis is still important in determining the outcome.

A constraint on the orientation of the post-impact angular momentum vector translates into a constraint on the inclination of the pre-impact orbit.  Since a planet is a small target on the scale of the Solar System, we can consider the plane perpendicular to the relative velocity to be analogous to a dart board or a clock face.  Even for an impactor with a very small pre-impact inclination it is possible for the strike to occur at the 12 O'clock or 6 O'clock positions at the top or bottom of the board (Fig.~\ref{fig:obliquity}b), and we expect the distribution of strike locations to be roughly uniform around the circle.  Clearly if all of the post-impact angular momentum is derived from the impact then an impact that strikes at the top or bottom will result in the post-impact angular momentum being oriented at 90$^{\circ}$ to the target orbit.  The smallest obliquity will occur if the impact strikes exactly in the plane of the target orbit, at the 9 O'clock or 3 O'clock positions of our planetary dart board.  In this case the obliquity will be equal to the inclination of the pre-impact orbit of the impactor (Fig.~\ref{fig:obliquity}c).  The obliquity will only be zero if the impactor and target orbits are co-planar (Fig.~\ref{fig:obliquity}a).  Note that in the extremely unlikely case of a head-on impact ($\theta=0$) the post-impact spin will be zero and thus the obliquity undefined.

Today the orbit of the Moon is inclined at 5.1$^{\circ}$ to the ecliptic, while Earth has an obliquity of 23.4$^{\circ}$ and the mutual obliquity between the spin axis of Earth and the orbit of the Moon varies between 18 and 28$^{\circ}$ due to the effects of precession and nutation.  The orbit of the Moon contains the majority ($>$80~per~cent) of the total angular momentum of the Earth-Moon system.  Over time tidal evolution has resulted in a transfer of angular momentum from the spin of Earth to the lunar orbit.  Earlier in this evolution when the Moon was closer to Earth the inclination of the lunar orbit was larger, while the obliquity of Earth and the mutual obliquity were smaller, with all three being probably around 10$^{\circ}$ initially \citep[e.g.][]{touma1994}.

If we assume that both the target and impactor are non-rotating initially, such that all of the angular momentum of the post-impact system is imparted in the impact, then we can use the initial obliquity of $\sim$10$^{\circ}$ of the Earth-Moon system to constrain the inclination of the impactor pre-impact orbit to be less than 10$^{\circ}$.  This does not result in any reduction in the parameter space for the Canonical and \citet{canup2012} models, but does exclude some low eccentricity pre-impact orbits for the \citet{reufer2012} models.

The \citet{cuk2012} scenario is more complicated, because here the pre-impact spin of the target is very fast and contributes the majority of the final angular momentum of the system.  While they only examined impacts in which the impactor strikes in the equatorial plane of the target their results show a distinct preference for retrograde impacts over prograde or head on impacts.  The study of \citet{canup2008} also showed that impacts in which the equatorial plane of the target is orthogonal to the impact lie roughly halfway between prograde and retrograde impacts in outcome. So we can expect that the preference for retrograde impacts over prograde impacts of \citet{cuk2012} will extend to a preference over orthogonal impacts.  As such, while the constraints we can place on the pre-impact inclination are not as strong as for the other scenarios, we can still exclude significant regions of the large range of inclinations allowed in Fig.~\ref{fig:EMCSaei}.  For example excluding inclinations over 20$^{\circ}$ (thus cutting Fig.~\ref{fig:EMCSaei} along the cyan band) allows us to exclude $\sim$91~per~cent of the orbits in the upper panel of Fig.~\ref{fig:EMCSaei} and $\sim$97~per~cent of the orbits in the lower panel of Fig.~\ref{fig:EMCSaei}.  Any exclusion of higher inclination orbits significantly increases the minimum eccentricity of the pre-impact orbit and increases the proportion of the solution space that is encompassed by the tail of Jupiter-crossing orbits.

For the case of the \citet{cuk2012} scenario it is also worth looking again at the very high relative velocities required.  In a population of self-stirred bodies, the velocity dispersion is roughly equal to the escape velocity, and so a random velocity of twice the escape velocity is unlikely.  This is supported by \citetalias{quarles2015}, who consider the distribution of the impacts in their simulations as a function of the impact velocity and find that there are very few impacts with velocities as high as 2~$v_{\rm esc}$, and almost no successful cases with impact velocities above 1.6~$v_{\rm esc}$.  Broader studies of terrestrial planet formation \citep[e.g.][]{obrien2006, quintana2016} similarly find that impacts at high relative velocities are unlikely, especially for embryo-embryo impacts.  

This low likelihood of high velocity impacts only applies to projectiles originating from within the inner Solar System, however. A Theia originating in the outer Solar System would necessarily have a higher typical impact velocity, due to the high eccentricity required to cross the Earth.  Dynamical studies of terrestrial planet formation do not typically include sources in the outer Solar System, and so it is difficult to judge the relative probabilities of a high velocity giant impact from within the inner Solar System versus an impact by an outer Solar System body during terrestrial planet formation.  If a high velocity impact scenario like that of \citet{cuk2012} becomes the favoured scenario for the formation of the Moon, then the possibility of an outer Solar System origin is of particular interest as it is likely that Theia would have been rich in volatiles including water.

\subsection{Discussion of Moon-formation scenarios}
\label{sec:moon:discuss}

From our analysis of the allowed distributions of pre-impact orbits we can divide the Moon-formation scenarios into two broad categories.  The Canonical, \citet{canup2012} and \citet{reufer2012} scenarios are all relatively low impact velocity events.  The \citet{cuk2012} scenario and the preferred impacts of \citet{rufu2017} occupy a second category as high velocity events.  The allowed pre-impact orbits for low-velocity events place the origin of Theia firmly within the inner Solar System, while for the high impact velocity events, outer Solar System origins are allowed.

\citet{quarles2015} show that the number of Earth-impactors originating from within the inner Solar System at the high velocities required for the \citet{cuk2012} scenario is very low, and the number of successful impacts is even lower.  They also find that within the inner Solar System an impactor that satisfied the impact conditions of the \citet{cuk2012} scenario would have more likely originated beyond the orbit of Mars, where the influence of Jupiter is stronger.  This is in accordance with our suggestion that the impactor in this high velocity scenario could have originated in the outer Solar System, for which a high impact velocity is guaranteed.  

While it is easier to obtain such a high impact velocity through a scattering encounter with Jupiter than through fortuitous multiple scattering within the inner Solar System, this is still an unlikely scenario.  Due to the high escape velocity of Jupiter relative to the Keplerian orbital speed, objects that undergo encounters with Jupiter are likely to be ejected from the Solar System \citep[e.g.][]{wyatt2017}.  This applies whether the object being scattered originates from the inner or outer Solar System.  Scattering by Jupiter is a necessary but not sufficient condition; the potential impactor must also be scattered onto a trajectory that crosses the inner planets, and it must then impact the desired terrestrial planet before it is ejected from the Solar System by a subsequent encounter with Jupiter.

A dynamical pathway that requires scattering with Jupiter prior to impact with the proto-Earth also requires tighter constraints on the inclination of the pre-impact orbit than from obliquity considerations since the orbits of Earth and Jupiter are well aligned (1.3$^{\circ}$ today).  The inclination of the pre-impact orbit must be comparably low for it to be able to intersect both planets.

In their simulations of the debris ejected by the Moon-forming impact, which we can take as a rough proxy for the behaviour of bodies in the inner Solar System, \citet{jackson2012} find that 2804 particles out of a total of 36000 are ejected from the Solar System over 10~Myr of simulated time, compared with 27 that hit the Sun, both outcomes that are dominated by scattering with Jupiter.  As such we might expect to need tens of potential impactors to scatter off Jupiter and be ejected for every one that is scattered onto an impact trajectory, and that is without consideration for whether that impact trajectory would have the correct relative velocity.  By comparison the terrestrial planets all lie in the `accreted' regime in \citet{wyatt2017}, indicating the ultimate fate of a body scattered by the terrestrial planets is most likely to be accretion.

We thus conclude that the \citet{cuk2012} scenario is a considerably less likely event than any of the lower impact velocity scenarios. Similarly, this conclusion applies to the \citet{rufu2017} scenario, in which each individual impact that generates a disk primarily composed of Earth material is a less likely event than any single lower-velocity Moon-formation scenario. Furthermore, of the moderate-to-high velocity scenarios of \citet{rufu2017}, $V_{\rm{imp}} \sim 2V_{\rm{esc}}$, which allow for predominantly Earth material to be distributed into the disk, these collisions occur primarily in head-on configurations. This impact configuration is at the tail end of the impact angle distribution for a giant impact, and is not as likely to occur as more probable configurations, at impact angles near 45$^\circ$, which contribute upwards of 30-60\% of impactor material to the proto-Lunar disk in their models, causing it to become isotopically distinct from Earth.

Our prediction, that Theia is unlikely to have originated from inside around 0.7~AU, is in agreement with the detailed investigation of \citetalias{quarles2015}. But more distant bodies are allowed; in particular, the water-rich impactor suggested by \citet{reufer2012} is consistent with the range of pre-impact orbits allowed given the relative velocity specified in their scenario.  Determining the relative likelihoods of the three low impact velocity scenarios, including hit-and-run, requires more detailed investigation with $N$-body simulations, like that of \citetalias{quarles2015}, who suggest that the Canonical and \citet{canup2012} scenarios are the most intrinsically likely scenarios due to their impact velocity and impact angle.

One notable factor that we predict but that is missing in the $N$-body study of \citetalias{quarles2015} is the pre-impact inclination.  Aside from a small subset with initial inclinations of 0.66$^{\circ}$ for Theia the simulations of \citetalias{quarles2015} all begin with Theia co-planar to Earth.  Given the non-zero inclinations we predict for the pre-impact orbit and the close match between the edge of the region of successes and pseudo-successes found by \citetalias{quarles2015} and our pre-impact orbit distribution, the effect of non-zero initial inclinations would be an interesting avenue to investigate in future studies.  A non-zero initial inclination for Theia would likely affect the final value of the angular momentum deficit for the terrestrial planet system.  Similarly, on the basis of our pre-impact orbit prediction and the trend in the results of \citetalias{quarles2015} there are likely initial conditions with semi-major axes in the range 1.1-1.2~AU and eccentricities greater than 0.1 that would result in success or pseudo-success.

\section{Mercury formation}
\label{sec:mercury}

We now turn to other giant impact scenarios. Of the four terrestrial planets, Mercury stands out as the most peculiar.  Unlike the rest of the terrestrial planets the composition of Mercury is dominated by its iron core, which comprises around 70~per~cent of the mass of the planet \citep{hauck2013}.  This is far higher than the close to chondritic iron content of the other terrestrial planets, which are also (aside from the Moon) significantly more volatile-rich than Mercury. This strongly suggests that Mercury underwent a unique formation process that caused it to deviate from the path followed by the rest of the terrestrial planets.

There are two principle classes of proposed methods of achieving the iron enrichment of Mercury: processes that alter the structure of the proto-planetary disk such that Mercury forms from a population of iron-rich dust or planetesimals, and processes that alter the structure of Mercury after it has formed to remove the bulk of the silicate mantle.  In the class of adjustments to the proto-planetary disk are suggestions that variation in the condensation of minerals in the inner regions of solar nebula would lead to iron-rich conditions \citep[e.g.][]{ebel2011}, and mechanisms to induce radial separation between iron and silicate dust grains/planetesimals \citep{weidenschilling1978}, including the recent suggestion of photophoresis \citep[e.g.][]{wurm2012}.  Removal of silicate mantle from a planetary mass proto-Mercury is mediated by collisions, either sand-blasting by many small impacts \citep[e.g.][]{svetsov2011} or by one or more giant impacts \citep[e.g.][]{benz2007, asphaug2014b}.  Here we will examine the giant-impact Mercury formation scenarios of \citet{benz2007} and \citet{asphaug2014b}.

\subsection{Proto-Mercury as the target}
\label{sec:mercury:benz}

In \citet{benz2007} the collisional removal of mantle material from proto-Mercury is achieved by impacting it at high speed with a smaller body, resulting in a net loss of mass from the proto-Mercury, primarily by the ejection of mantle silicate materials. The target proto-Mercury has a mass 2.25 times the present mass of Mercury, and in their successful scenarios the impactor has a mass of 0.375-0.45 Mercury masses, with relative velocities of 28-30~km~s$^{-1}$, around 6 times the escape velocity of the proto-Mercury.  They also find one marginally successful case at 20~km~s$^{-1}$ for a directly head-on impact.  While Mercury formation by such a process is sometimes thought of as a `mantle stripping' event, the final Mercury is less than half the mass of the target, placing this collision in the catastrophic disruption regime \citep[e.g.][]{leinhardt2012}. The final Mercury produced in this case is composed of the gravitationally re-accumulated remnants of the disrupted core of proto-Mercury plus a lesser amount of mantle material.

\begin{figure}
 \includegraphics[width=\columnwidth]{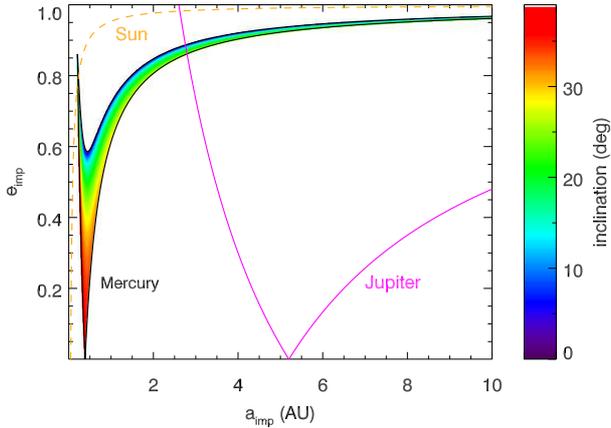}
 \caption{Allowed semi-major axis -- eccentricity -- inclination distribution for impactors striking a proto-Mercury at $v_{\rm rel}=30$~km~s$^{-1}$ per the \citet{benz2007} model. We assume that proto-Mercury has a circular orbit at 0.387~AU.  Orbits above the magenta line are Jupiter crossing, while those above the orange line are Sun-grazing (passing within 10~$R_{\oplus}$ of the Sun).}
 \label{fig:mercurytar}
\end{figure}

As this is a very violent impact, and the final product contains only a fraction of the total mass of the colliding bodies, the assumption that the orbit of the target will be approximately the same as that of the present day final body is likely to break down.  As such, rather than assuming that the proto-Mercury follows the same orbit as the present day Mercury, here we will assume that it follows a circular orbit at the present day semi-major axis of Mercury.  While it is not clear that this would adequately approximate the true orbit of proto-Mercury, it does seem likely that proto-Mercury would have had a less excited orbit than the remnant of the giant impact that is the present day Mercury. At a semi-major axis of 0.387~AU the relative velocity of 30~km~s$^{-1}$ translates into $v_{\rm rel}/v_{\rm k}$=0.63.  We show the resulting distribution of allowed impactor orbits in Fig.~\ref{fig:mercurytar}.

\subsection{Proto-Mercury as the impactor}
\label{sec:mercury:asphaug2014b}
 
 There are two main problems with the catastrophic disruption scenario where Mercury is the target. One, after the collision the newly formed Mercury would be orbiting through a massive torus of silicate debris, its catastrophically disrupted mantle, most of which continues orbiting the Sun. One must ensure that Mercury does not accrete too much of this material or else the core fraction will become too low. Two, the event is more intensely energetic than the formation of Earth's Moon, thus Mercury would be expected to be at least as devolatilized as the Moon, when in fact it is comparatively rich in volatiles \citep{paige2013} and indeed Earth- and Mars-like in semi-volatiles \citep{peplowski2011radioactive}.
 
 \citet{asphaug2014b} propose an alternative giant impact scenario for the formation of Mercury.  Rather than the proto-Mercury being struck by a smaller impacting body that results in its disruption and the loss of its mantle, proto-Mercury is itself the impactor and collides in a hit-and-run impact with a larger body, either proto-Venus or proto-Earth. This hit-and-run scenario solves the problem of mantle re-accumulation by Mercury, since proto-Venus or proto-Earth dominate the accretion of most of the mantle material liberated from proto-Mercury by the giant impact.  Of course, the hit-and-run scenario introduces a new problem in that the newly formed Mercury will still be on an orbit which crosses one or both of the orbits of Venus and Earth and is reliant on other dynamical interactions perturbing it away from further collisions.  This can happen \citep{chambers2013}, and although most such Mercuries would be accreted by Venus or Earth, those are gone (now part of Venus or Earth), so \citet{asphaug2014b} argue that the probabilities of survival of the core of a single hit and run proto-Mercury is consistent.
 
 While the hit-and-run scenario is conceptually very different from the \citet{benz2007} scenario, the relative velocity required in the two cases is actually quite similar.  The best fit found by \citet{asphaug2014b} for producing Mercury in a single hit-and-run event uses a relative velocity of around 30~km~s$^{-1}$, and they find that all successful single hit-and-run scenarios have relative velocities of $>$25~km~s$^{-1}$.  Since the orbital velocity at Venus or Earth is lower than at the semi-major axis of Mercury this actually results in a higher $v_{\rm rel}/v_{\rm k}$ of $>$0.71 if the target is Venus or $>$0.84 if the target is Earth, assuming that the proto-Venus and proto-Earth in this scenario have orbits close to their present day ones, as is likely since they are near their present day masses (0.85~$M_{\oplus}$).
 
 \begin{figure}
  \includegraphics[width=\columnwidth]{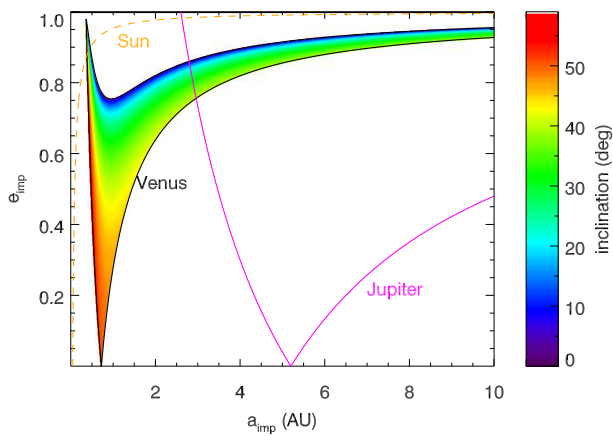}
  \caption{Allowed pre-impact orbits for proto-Mercury impacting Venus at 3~$v_{\rm esc}{(\rm Venus})$ as suggested by \citet{asphaug2014b} for a single hit-and-run collision. Orbits above the magenta line are Jupiter crossing, while those above the orange line are Sun-grazing.}
  \label{fig:mercuryimp}
 \end{figure}

 We show the orbital distribution for the pre-impact proto-Mercury for an impact with Venus at 3~$v_{\rm esc}$ ($v_{\rm rel}$ = 29.3~km~s$^{-1}$, $v_{\rm rel}/v_{\rm k} = 0.84$) in Fig.~\ref{fig:mercuryimp}.  As we can see the distribution is broader than that in Fig.~\ref{fig:mercurytar} due to the higher $v_{\rm rel}/v_{\rm k}$ at Venus.  If the target is Earth then the distribution becomes even more extreme as $v_{\rm k}$ drops further and the escape velocity of Earth is slightly higher, such that a 3~$v_{\rm esc}$ impact at Earth requires a relative velocity above the orbital velocity of Earth, $v_{\rm rel}/v_{\rm k} = 1.06$.  This leads to retrograde orbits for proto-Mercury in the upper left of the distribution.
 
 Though \citet{asphaug2014b} demonstrated a successful scenario for the production of Mercury in a single hit-and-run impact with either proto-Earth or proto-Venus, their preferred scenario is for the proto-Mercury to undergo two or three hit-and-run collisions with proto-Earth and/or proto-Venus, losing a fraction of its mantle in each collision and progressively increasing the core-mantle mass ratio. They argue for multiple hit-and-runs statistically, on the grounds that a hit-and-run survivor is likely to have survived more than one such collision, and geochemically, on the grounds that two or more individual impacts would each be less violent, with lower relative velocities than the single hit-and-run scenario, thus preserving more volatiles.  \citet{asphaug2014b} do not identify a specific impact sequence, but generalizing from previous hit-and-run results \citep{asphaug2010} they argue that 2-3 impacts at a relative velocity roughly equal to the escape velocity of the target (proto-Earth or proto-Venus) would be sufficient to increase the core-mantle ratio of an initially chondritic proto-Mercury to what is observed today.
 
 \begin{figure}
  \includegraphics[width=\columnwidth]{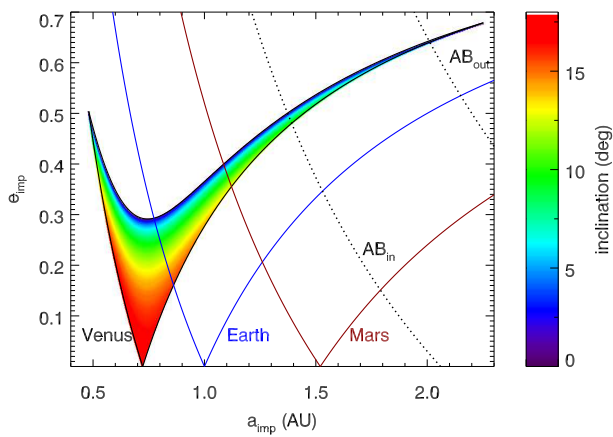}
  \caption{Allowed pre-impact orbits for proto-Mercury impacting Venus at $v_{\rm rel}=v_{\rm esc}(\rm{Venus})$, as suggested by \citet{asphaug2014b} for producing Mercury through a sequence of 2-3 hit-and-run impacts.  Orbits above the blue line are Earth crossing, while those above the dark red line are Mars crossing.  The black dotted lines mark the approximate boundaries at which the impactor would be crossing the inner and outer edges of the asteroid belt.}
  \label{fig:mercuryimp2}
 \end{figure}
 
 In fig.~\ref{fig:mercuryimp2} we show the allowed distribution of pre-impact orbits for proto-Mercury for an impact with Venus at $v_{\rm rel}=v_{\rm esc}(\rm{Venus})$.  In contrast with both the single hit-and-run model and the \citet{benz2007} model the allowed distribution of pre-impact orbits is now much smaller and fully contained within the terrestrial planet system with no Jupiter crossing region.  If the target is changed to Earth then a small region of Jupiter crossing orbits for proto-Mercury is allowed.
 
 \subsection{Discussion of Mercury-formation scenarios}
 \label{sec:mercury:discuss}
 
 The relative velocities required for both classes of single impact scenarios, those with proto-Mercury as the target and those with proto-Mercury as the impactor, are similar at around 30~km~s$^{-1}$ in both cases. This is also similar to the relative velocity in the Moon-formation scenario of \citet{cuk2012} (Section~\ref{sec:moon:cuk}).  As such, these scenarios both result in similarly broad distributions of allowed orbits for the impactor and fall into the same high velocity category.  The distribution for the \citet{benz2007} proto-Mercury-as-target scenario is slightly narrower than that for the \citet{asphaug2014b} proto-Mercury-as-impactor scenario, which is in turn slightly narrower than that for the \citet{cuk2012} Moon-formation scenario, since the higher orbital velocities closer to the Sun translate into lower $v_{\rm rel}/v_{\rm k}$ for the same $v_{\rm rel}$.
 
 Occupying the same high relative velocity category as the \citet{cuk2012} Moon-formation scenario, many of the same conclusions regarding the relative likelihood of that scenario also apply to the single impact Mercury formation scenarios.  Achieving the relative velocities necessary without the influence of Jupiter is difficult, and outer Solar System origins for the impactor are possible.  As previously, if the impactor originates in the outer Solar System this places tight constraints on the pre-impact inclination to allow for both scattering with Jupiter and collision with the relevant terrestrial planet.
 
 If the impactor did indeed originate in the outer Solar System then this has different implications for the model of proto-Mercury as the target versus proto-Mercury as the impactor.  For the proto-Mercury-as-target model, an origin of the impactor in the outer Solar System may be beneficial in the same way as for the \citet{cuk2012} and \citet{reufer2012} Moon-formation models.  An outer Solar System impactor would likely be water-rich, which would decrease the mass of silicate material in the impactor and reduce the problem of avoiding re-accretion of the silicate material.  While a water-rich impactor could potentially introduce a new problem of avoiding the re-accretion of water, it will be vapourised and dispersed by the energetic giant impact leaving behind a situation analogous to the formation of Earth's Moon. In addition, it would be difficult for the low mass Mercury to retain substantial amounts of water in its thermal and radiation environment in the long term, albeit that small amounts can be retained in shadowed regions at the poles \citep{lawrence2013, paige2013}. 
 
 For the single hit-and-run origin of Mercury, however, an origin for the impactor (in this case proto-Mercury itself) in the outer Solar System, which would thus likely be water-rich, presents a more problematic issue.  In a water-rich proto-Mercury the iron fraction of the total body would be smaller, presuming a chondritic silicate-iron ratio with an addition of water.  This increases the required total mass of the body if it is to contain the same mass of iron.  In addition, for a fully differentiated body with a water ice outer mantle, a rocky inner mantle and an iron core, the silicates and iron would now be more strongly concentrated at the centre of the body, with consequently higher (negative) total binding energy, and it would thus require more energy to remove the silicates overlying the iron.  The increased concentration of the iron toward the centre of the body, combined with the larger mass required to obtain the same total mass of iron, both suggest that the impact would need to be more violent than originally proposed by \citet{asphaug2014b}.  
 
 While such an energetic hit-and-run would be plausible for a proto-Mercury scattered by Jupiter, it raises questions about the consequences for Earth or Venus if they were struck with such intensity.  For example, mantle stripping of the target (Earth or Venus) could become significant, as could the loss of volatile elements from a very strongly shock heated and highly deformed target planet. Conversely, as it is the target that accumulates most of the mantle ejected from proto-Mercury in the hit and run scenario, the hit and run by a water-rich impactor would additionally contribute a late complement of outer Solar System water onto the target, in addition to the stripped mantle silicates. These complexities do not rule the scenario out -- Venus and Earth evolved quite differently for some reason -- but raises the importance of understanding the implications for the target of the hit-and-run scenario.
 
 In the multiple hit-and-run Mercury-as-impactor scenario proposed by \citet{asphaug2014b}, while proto-Mercury could have been water-rich if it originated in the asteroid belt, this is not a necessary condition.  Furthermore, a relative velocity of around the escape velocity of Earth or Venus is readily expected for an impact involving a body that has been scattered by the terrestrial planets, and indeed Jupiter crossing orbits are not allowed if Venus is the target.
 
 While a scenario involving a chain of impacts might seem at first glance unlikely, the individual impacts are fairly representative of the impacts one would expect during terrestrial planet formation.  \citet{kokubo2010} and \citet{chambers2013} find that around half of all impacts are hit-and-run, and so a chain of several hit-and-runs is not especially unlikely, for a given unaccreted embryo. According to \citet{asphaug2014b} the accreted embryos are biased to include all of the slow, direct hits, causing the unaccreted embryos to be biased to include an increasing fraction of hit and run survivors. Indeed in one of the handful of cases in \citet{chambers2013} they find a Mercury-sized survivor of four successive hit-and-run collisions.  
 
 The single impact scenarios however require rather extreme events.  Following the same arguments as for the \citet{cuk2012} Moon-formation model in Section~\ref{sec:moon:discuss} the single impact scenarios are thus unlikely and on the basis of our work we suggest that the multiple hit-and-run scenario of \citet{asphaug2014b} is statistically more favourable, although requiring more detailed study in terms of solar system dynamics and collision geophysics.
 
 \section{The Borealis basin and eccentric targets}
\label{sec:borealis}
 
A dominant characteristic of Mars is the hemispheric crustal dichotomy, with the northern hemisphere lying on average 5~km lower than the southern hemisphere \citep[e.g.][]{smith1999}.  It has long been recognised that a massive impact might be able to reproduce the hemispheric dichotomy \citep[e.g][]{wilhelms1984}.  More recently, improvements in our knowledge of the gravity field of Mars \citep{andrews-hanna2008} and numerical modelling of giant impacts into Mars \citep{marinova2008, nimmo2008, marinova2011} has shown that a giant impact indeed seems to well reproduce the hemispheric dichotomy and fits well with other features of Mars, such as its magnetic field.  Additionally, \citet{citron2015} demonstrate that a Borealis basin-forming collision places enough mass into a circum-Mars disc to produce a protolunar disk from which the Martian moons Phobos and Deimos could derive.  Under this scenario the hemispheric dichotomy is then the scar of a giant impact involving a planetesimal around twice the mass of Ceres.

\begin{figure}
 \includegraphics[width=0.95\columnwidth]{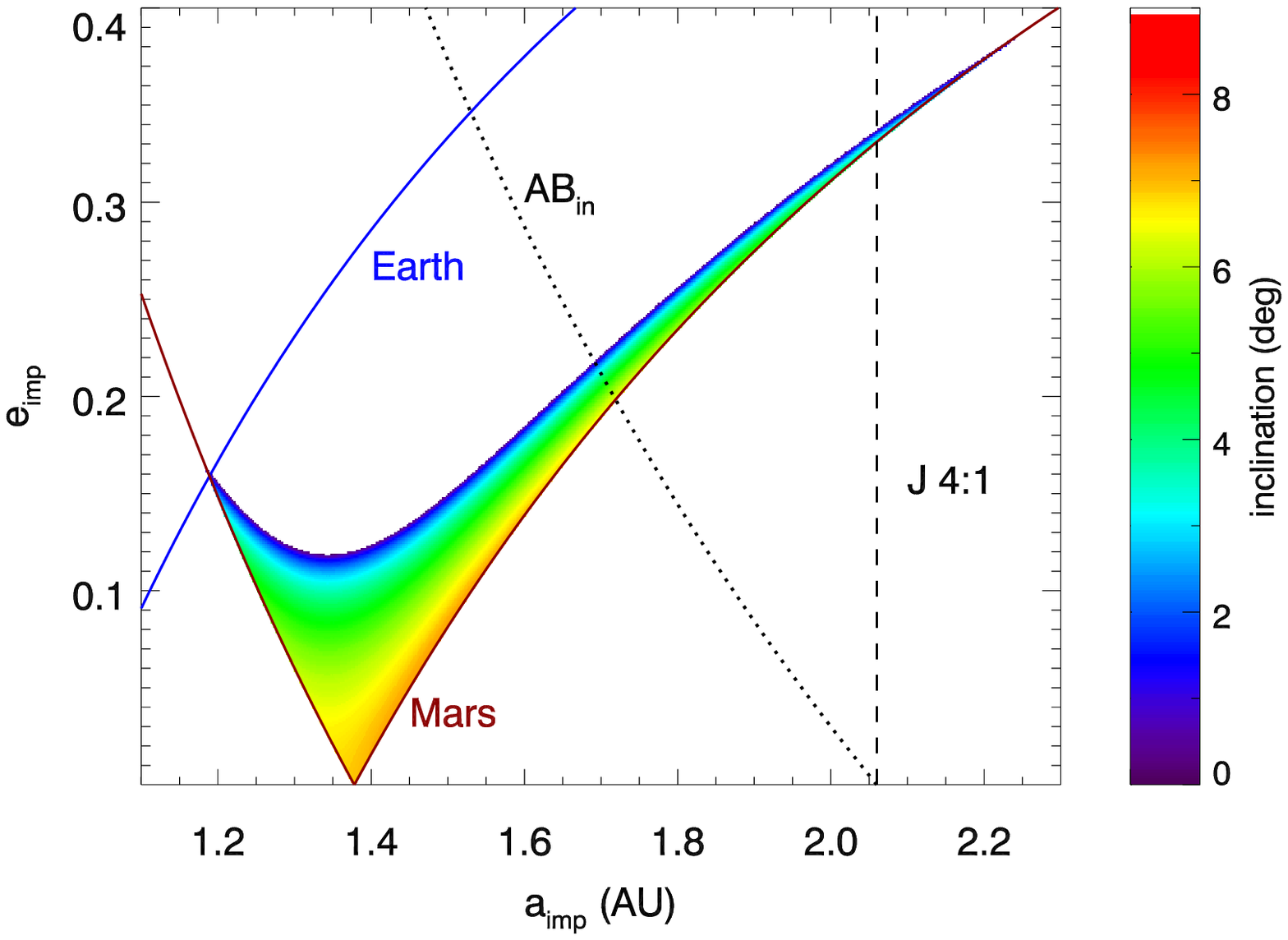}
 \includegraphics[width=0.95\columnwidth]{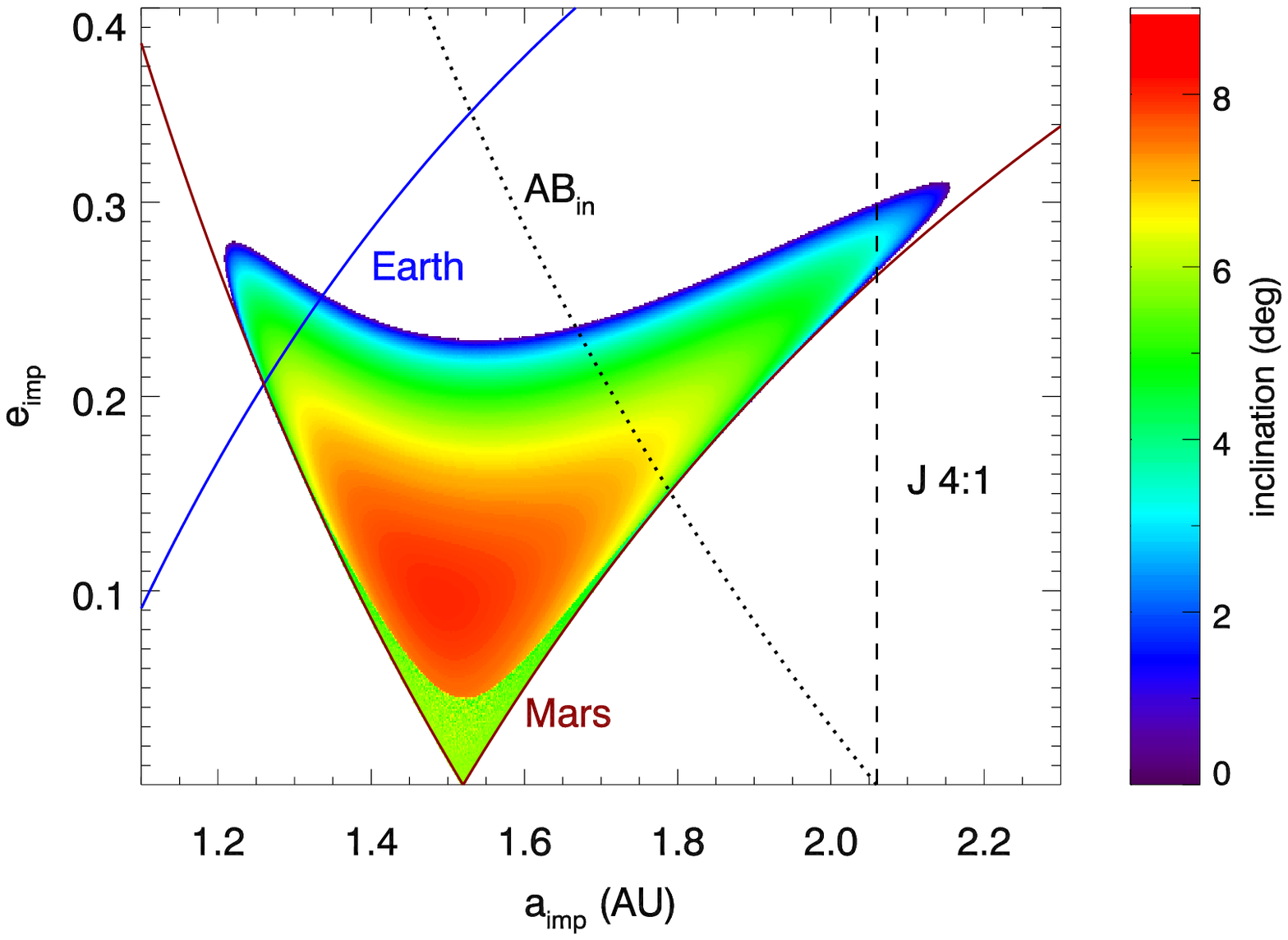}
 \includegraphics[width=0.95\columnwidth]{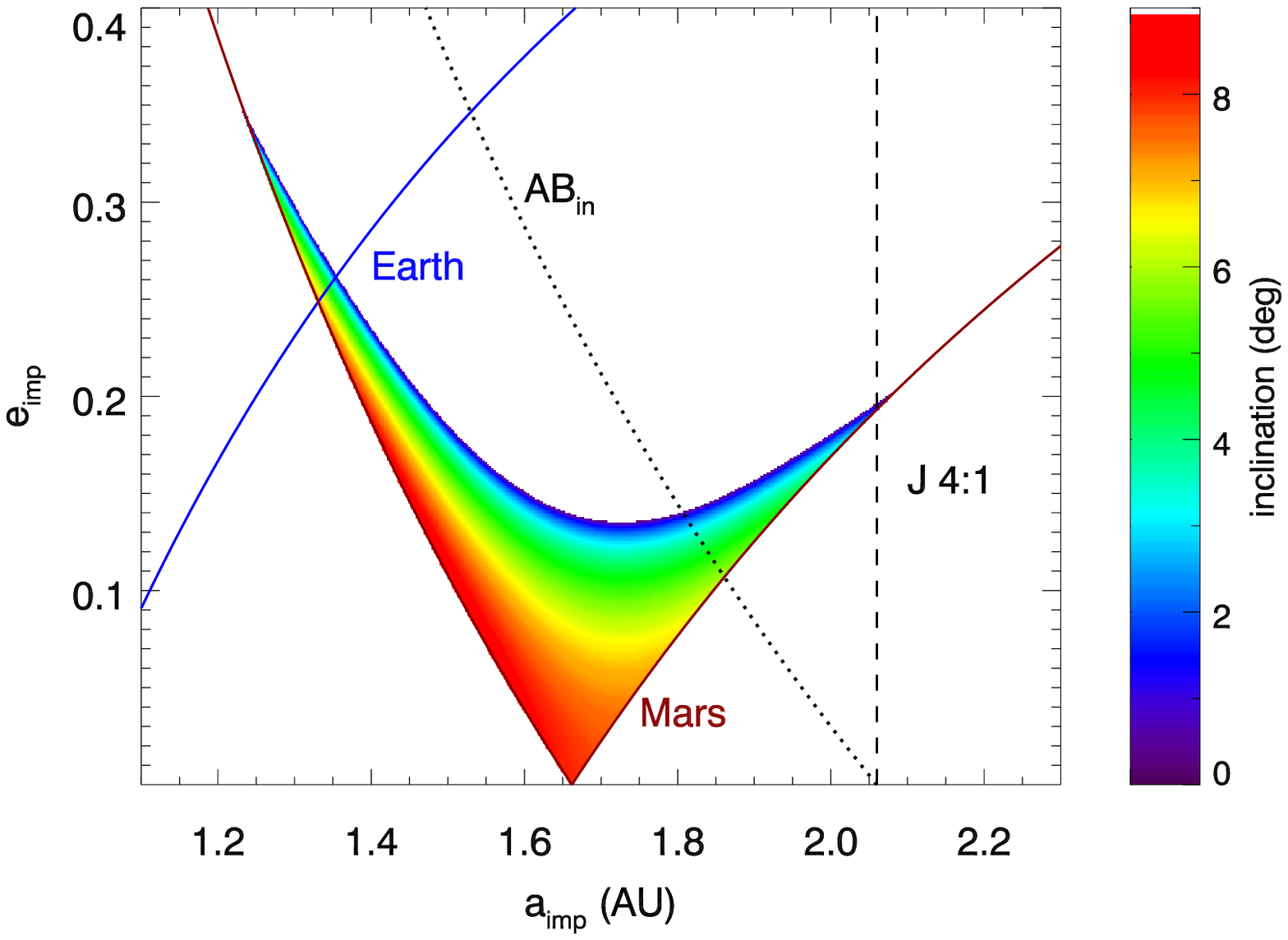}
 \caption{Allowed semi-major axis -- eccentricity -- inclination distributions for the pre-impact orbit of the Borealis impactor (assuming the best-fit parameters of \citealt{marinova2011}) for an impact occurring when Mars is at pericentre (top), at an orbital distance of 1.52~AU (equal to the Martian semi-major axis, middle), and at apocentre (bottom).  Orbits above the blue line are Earth crossing, while those above the dotted line have apocentres beyond 2.06~AU, the inner edge of the asteroid belt.  The black dashed line indicates the 4:1 mean motion resonance with Jupiter.}
 \label{fig:borealis}
\end{figure}

The Borealis basin impact also represents a useful demonstration case, since the eccentricity of the Martian orbit is significant and so must be taken into account when examining the potential pre-impact orbit of the Borealis impactor.  As we can see from the equations in Section~\ref{sec:method} the key difference is that when the target has an eccentric orbit the range of impactor orbital parameters that will produce the correct relative velocity now depends on the true anomaly of the target at the time of impact, since the velocity of the target varies it moves around its orbit.  Since we cannot easily represent the variation in the allowed impactor orbit distribution as a function of the true anomaly of Mars at impact over the whole range of true anomaly, we instead chose three representative locations, when Mars is at apocentre, pericentre and when the Sun-Mars separation is equal to the Martian semi-major axis ($r=a$).  Apocentre and pericentre represent the extremes, when Mars is moving slowest and fastest respectively, while $r=a$ provides a useful middle ground, since at this location the velocity of Mars is equal to that of a body on a circular orbit at $a_{\rm Mars}$.  These three cases are shown in the top, middle and bottom panels of Fig.~\ref{fig:borealis}, for which we use the best-fit scenario of \citet{marinova2011} in which a roughly 2000~km body impacts Mars in a glancing collision at around 1.2~$v_{\rm esc}$ or $v_{\rm imp}$=6~km~s$^{-1}$.  For the Martian escape velocity of 5.027~km~s$^{-1}$ this corresponds to $v_{\rm rel}/v_{\rm k}$=0.14, or $v_{\rm rel}$=3.33~km~s$^{-1}$.

As we would expect from a low velocity impact the pre-impact orbits for the Borealis impactor lie close to Mars.  Regardless of where on the Martian orbit the impact takes place a significant region of impactor orbits that cross into the asteroid belt are allowed, due to the proximity of the asteroid belt to Mars.  In contrast Earth crossing orbits are allowed only when the impact takes place away from the pericentre of the Martian orbit.  This may seem counter-intuitive since Mars is closest to Earth at pericentre, but is due to the faster orbital speed of Mars at this point.  Since the allowed range for pre-impact orbits that also cross the orbit of Earth is small, this suggests that it is unlikely that the impactor would have been passed outwards by the other terrestrial planets.  As such we recommend that any future detailed $N$-body examination of the origin of the impactor proposed for the Borealis basin should focus on the range from 1.2-2.2 AU, near Mars.
 
 \section{Other Solar System giant impacts}
 \label{sec:otherimpacts}
 The terrestrial planets are not the only places where there are features that are best explained by a giant impact.  In the outer reaches of the Solar System a giant impact has also been proposed as the origin for the Pluto-Charon system \citep[e.g.][]{canup2005,canup2011,stern2006}.  The successful simulations of \citet{canup2011} had low impact velocities of $<$1.2~$v_{\rm esc}$, however translating this into predictions for the pre-impact orbit of the impactor is difficult since the orbit of proto-Pluto at the time of impact is highly uncertain.  Today Pluto-Charon is in a 3:2 mean-motion resonance with Neptune, a result of Neptune's outward migration into the Kuiper belt. It is likely that the formation of Charon occurred before or during the migration of Neptune as the primordial Kuiper belt would have been much more massive with many more potential impactors \citep[e.g.][]{tsiganis2005}.  As such, the orbit of proto-Pluto at the time of impact was likely not the same as Pluto-Charon today.  It has however worth noting that whatever the orbit of proto-Pluto at the time of impact the low impact velocity effectively requires that the impactor must have originated from the same population as proto-Pluto.  If the impact occurred after capture into the present, eccentric, orbit of Pluto-Charon then impactors with eccentricities below around 0.1 can be excluded, indicating that the impactor would be another resonant or scattered disk object.  Conversely if the impact occurred before capture into resonance then the impactor would have been a low eccentricity object from nearby.  If future constraints, for example on the timing of the Pluto-Charon forming impact, allow us to estimate what the orbit of proto-Pluto would have been at the time of impact, then we would be able to distinguish between these possibilities.
 
 Smaller scale events are evident in the creation of the Hirayama asteroid families \citep[e.g.][]{durda2007}.   In particular, the Vesta family is believed to be traceable to the formation of the massive Rheasilva basin $\la$1~Gyr ago \citep{binzel1997}, which dominates the southern hemisphere of Vesta and which models suggest was caused by an impact around 66~km in diameter \citep{jutzi2013}, large compared with the 525~km mean diameter of Vesta itself.  Similarly in the outer Solar System the dwarf planet Haumea also has a collisional family associated with it \citep{brown2007}, which was presumably also created in a large impact, perhaps with a precursor of the current two satellites of Haumea \citep{schlichting2009,cuk2013}.  It has also long been recognised that a large impact could explain the large obliquity of Uranus \citep[e.g.][]{safronov1972,parisi1997}.
 
\section{Conclusions}
\label{sec:conclusions}

The impact velocity is one of the key parameters in any giant impact scenario and we have demonstrated how this can be used to provide constraints on the orbit of the impactor immediately prior to the collision.  These constraints are very fast to compute, taking only a few seconds on a laptop computer.  This allows us to quickly compare a wide range of giant impact scenarios in a uniform way and make broad judgements on the likelihood of different scenarios as well as narrow down the range of initial orbital parameters within which more detailed examinations with expensive $N$-body simulations should be targeted.  We apply this method to examine the currently proposed scenarios for giant impacts in the Solar System for the formation of the Moon, Mercury and the Borealis basin on Mars.  Any subsequent proposals for giant impact scenarios in the Solar System can then be compared in the same manner.

The \citet{cuk2012} scenario for the formation of the Moon, both current scenarios for the formation of Mercury through a single impact \citep{benz2007, asphaug2014b}, and the multiple impact Moon-formation scenario of \citet{rufu2017} all have similar distributions of pre-impact orbits and can be grouped together as high-velocity impacts.  For all of the scenarios in this category achieving the required impact velocity through scattering with the terrestrial planets is unlikely, but can be readily achieved by scattering with Jupiter.  Scattering of an object by Jupiter onto an impact trajectory with a terrestrial planet is still a low probability event however and so we disfavour all of the scenarios in this high impact velocity category.  If we rule out these high velocity scenarios, this means that in the case of Mercury, the only remaining scenario we consider favourable from a probabilistic perspective is the multiple hit-and-run proposal of \citet{asphaug2014b}.

While we cannot distinguish the three low-velocity scenarios for the formation of the Moon in terms of relative probability, we can predict the range of orbital parameters from which we expect the impactor to have originated.  Comparing these predictions with the detailed $N$-body study of \citet{quarles2015} demonstrates that our predicted parameter ranges do indeed encompass the origins of successful impactors.  As such we recommend that any future $N$-body examination of the origin of the Borealis basin impactor should focus on the range from 1.2-2.2~AU near Mars.  
 
\section*{Acknowledgements}

The authors gratefully acknowledge funding through NASA grant NNX16AI31G (Stop hitting yourself). The results reported herein benefited from collaborations and/or information exchange within NASA's Nexus for Exoplanet System Science (NExSS) research coordination network sponsored by NASA's Science Mission Directorate.  The authors are grateful to Billy Quarles for providing the data from Fig.~2d of \citetalias{quarles2015} for use in Fig.~\ref{fig:EMcanon-QL}.  APJ thanks Dan Tamayo for useful discussions.  The authors thank the anonymous referee for insightful suggestions that have improved the manuscript.

{\footnotesize
\bibliographystyle{mnras}
\bibliography{preimporbs}
}
 
 \label{lastpage}
\end{document}